\documentclass[twocolumn,showpacs,preprintnumbers,amsmath,amssymb]{revtex4}
\usepackage{graphicx}
\usepackage{dcolumn}
\usepackage{bm}
\usepackage{wasysym}
\usepackage{amsmath}
\usepackage{amssymb}

\newcommand{\aj}{AJ}

\newcommand{\mnras}{MNRAS}

\newcommand{\aap}{A\&A}

\newcommand{\jcap}{JCAP}

\newcommand{\mcut}{M_{cut}}
\newcommand{\cvir}{c_{vir}}
\newcommand{\rvir}{R_{vir}}
\newcommand{\mvir}{M_{vir}}
\newcommand{\ep}{\epsilon}
\newcommand{\Phiv}{\Phi_{vir}}

\newcommand{\mcl}{m_{cl}}
\newcommand{\rcl}{r_{cl}}

\begin{document}

\title{Can the dark matter annihilation signal be significantly boosted by substructures?}
\author{A. N. Baushev}
\affiliation{DESY, 15738 Zeuthen, Germany\\
 Institut f\"ur Physik und Astronomie, Universit\"at Potsdam, 14476
Potsdam-Golm, Germany\\
Departamento de Astronom\'ia, Universidad de Chile, Casilla 36-D, Correo Central, Santiago,
Chile\\}

\date{\today}

\begin{abstract}
A very general cosmological consideration suggests that, along with galactic dark matter halos,
much smaller dark matter structures may exist. These structures are usually called 'clumps', and
their mass extends to $10^{-6} M_\odot$ or even lower. The clumps should give the main contribution
into the signal of dark matter annihilation, provided that they have survived until the present
time. Recent observations favor a cored profile for low-mass astrophysical halos. We consider cored
clumps and show that they are significantly less firm than the standard NFW ones. In contrast to
the standard scenario, the cored clumps should have been completely destroyed inside $\sim
20$~{kpc} from the Milky Way center. The dwarf spheroidals should not contain any dark matter
clumps. On the other hand, even under the most pessimistic assumption about the clump structure,
the clumps should have survived in the Milky Way at a distance exceeding $50$~{kpc} from the
center, as well as in low-density cosmic structures. There they significantly boost the dark matter
annihilation. We show that at least $~70\%$ of the clumps endured the primordial structure
formation should still exist untouched in the present-day Universe.
\end{abstract}


\maketitle

\section{Introduction}
A very general cosmological consideration suggests that, along with large dark matter haloes,
corresponding to visible astronomical objects, like galaxies or galaxy clusters, much smaller dark
matter structures may exist. Indeed, the inflation inevitably leads to a Zel'dovich-Harrison-like
spectrum of the primordial cosmological perturbations, which is flat and contains perturbations of
any scale with approximately the same amplitude \citep{gorbrub2}. The perturbations with masses
lower than some cutoff value $\mcut$ may be destroyed in the early Universe by the free streaming.
If the dark matter (hereafter DM) is cold and non-interacting, which is the most popular hypothesis
now, $\mcut$ is mainly determined by the mass of dark matter particle $m_{dm}$. Roughly speaking,
$\mcut\sim M^3_{Pl}/m^2_{dm}$, where $M_{Pl}$ is the Planck mass. For $m_{dm}=100$~{GeV}, different
estimations of $\mcut$ vary from $10^{-12} M_\odot$ \citep{lowmassclump} to $10^{-6} M_\odot$
\citep{highmassclump}. We denote the clump virial mass and radius by $\mcl$, $\rcl$.

We consider the DM particles (hereafter DMPs) of mass $100$~{GeV} not by accident. If we suppose
that dark matter was in thermodynamical equilibrium in the early Universe, current DM abundance
allows estimating of the DM annihilation cross section $\langle \sigma v\rangle_a \simeq 3\cdot
10^{-26} \text{cm}^3/\text{s}$. This is a typical value for weak interactions, which implies that
DMPs may interact weakly. Of course, other explanations of the dark matter abundance are possible,
and the 'correct' cross section value can be just a coincidence. However, the weakly-interacting
massive particles (WIMPs) are one of the most popular dark matter candidate now. The WIMP mass in
realistic models typically exceeds $100$~{GeV}. A lighter WIMP is possible, but a theorist should
take special precautions to avoid tensions with the LHC results, since a light WIMP is usually
accompanied by respectively light charged particles in theories like supersymmetry. Thus the
existence of tiny dark matter halos (traditionally named 'clumps') is not a theoretical assumption,
but an inevitable property of many models of the cold dark matter, if DMP mass exceeds $\sim
1$~{MeV}.

The correct estimation of the dark matter clumpiness is of critical importance, especially for the
indirect dark matter detection. Indeed, the number of annihilations in a volume $dV$ in an interval
of time $dt$ is\footnote{We do not consider the situation when the dark matter particle is
identical to its antiparticle. The multiplier should be $1$ instead of $\frac12$ in this case.}
\begin{equation}
\dfrac12 \langle\sigma\upsilon\rangle\: \frac{\rho^2}{m^2_\chi} d\tau \, dV  \label{19b1}
\end{equation}
where $\langle\sigma\upsilon\rangle$ is the averaged multiplication of the DM annihilation
cross-section on the relative velocity and $m_\chi$ is the DMP mass.
 Thus the annihilation signal is proportional to the so-called boost
factor $C=\langle\rho^2\rangle/\langle\rho\rangle^2$, which takes into account the enhancement of
the annihilation signal by dark-matter substructures. In the standard cosmological scenarios the
smallest halos collapse the first, and therefore have the highest density contrast, making the main
contribution to the boost. We can define a characteristic radius $r_s$ of each halo where its
density profile gets steeper than $r^{-2}$, i.e., $d\log\rho/d\log r=-2$ at $r_s$. Then we may
introduce the halo concentration $c_{vir}\equiv r_s/R_{vir}$. In the case of NFW halo, this
definition coincides with the standard one.

The information about the smallest DM structures is poor at recent. First of all, astronomical
observations of the clumps are impossible, since their gravitational potential is too small to hold
baryon matter. The theoretical view on the clump structure and velocity distribution of the DM
particles is also rather uncertain. The clump mass is very small ($<10^{-6} M_\odot$ for neutralino
dark matter) comparing to the galaxy masses, and results and relationships obtained in standard
N-body simulations should be extrapolated on many orders of magnitude \citep{diemand2013}.
Moreover, as we will see below, the present-day number of clumps strongly depends on their
interaction with various baryon objects, like stars, galaxy disc etc. It makes a reliable
simulation of clump abundance very difficult.

The main fraction of the clumps formed in the early Universe has not survived to our time. There
are many mechanisms of clump demolition, and they can all be divided into two groups. A significant
part of clumps was tidally destroyed almost immediately after their occurrence by larger halos
during the hierarchical structure formation. We will imply the standard $\Lambda$CDM model
hereafter in this paper. In the framework of this scenario, the hierarchical clump destruction is
not sensitive to the clump density profile. Indeed, the spectrum of primordial perturbations is
flat in this case, and objects of different masses collapse almost simultaneously. The time
interval between the occurrence of the smallest clumps and more massive objects that capture the
clumps and destroy them tidally is smaller than the clump virilization time. Consequently, even if
present-day clumps have cusps, they were formed later. Therefore, we may use a simplified
estimation of the hierarchical clump destruction by calculating an energy gain per each tidal
interaction and the number of tidal interactions \citep{berezinsky2006}: a clump is considered to
be destroyed if its internal energy increase due to the tidal shocks exceeds its total energy
$|E|\sim GM/R$. This estimation leads to the $\propto dM/M^2$ mass spectrum of the clumps survived
the structure formation and suggests that only about $0.1 - 0.5\%$ of clumps endure the
hierarchical tides in the each logarithmic mass interval. N-body simulations \citep{diemand2005}
support the mass spectrum, but suggest a several times higher clump survival ratio. However, even
$0.1 - 0.5\%$ of clumps endured are sufficient to provide a very significant boosting
\citep{berezinsky2008}.

The second large group of clump destruction mechanisms can be named astrophysical. The clumps can
be disrupted by separate stars, galaxy stellar discs, tidal perturbations in the galactic
gravitational field, gas clouds etc. Contrary to the hierarchical destruction, this process goes on
up to now, and its efficiency is critically defined by the internal structure of the clumps. We
will consider these processes in the present  work and try to show that the clumps may be much less
persistent than it is generally believed.

\section{Energy structure of clumps}
It might appear at first sight that the difference between a cuspy and a cored profile is not that
large, especially if the core is small. However, this conclusion is incorrect, if we take into
account a possible velocity distribution anisotropy of DMPs. In order to illustrate it, we consider
the evolution of DMP energy distribution during the halo formation. The halo is formed from a
linear initial perturbation. The potential well of it was shallow. As a result, the initial total
energies $\ep=\frac{v^2}{2}+\phi$ of the particles forming the halo were close: under very general
assumptions, they lay in the interval $\ep\in[-1.5\Phiv;-\Phiv]$ (where
$\Phiv=G\dfrac{\mvir}{\rvir}$), i.e., they differed no more than $1.5$ times \citep{15}. The final
potential well is much deeper. The ratio of its depth (i.e., the central potential $\phi(0)$) to
the potential on the virial radius $-\Phiv$ is $\phi(0)/\Phiv\sim 30$ for the Milky Way
\citep{suchkov} and even larger for smaller objects, since they should be more concentrated.

It is convenient to characterize each particle, instead of its energy, by its apocenter distance
$r_0$, which is the largest distance that the particle can move away from the center. $r_0$ is
defined by the particle energy and the shape of the potential well. For instance, if the particles
had kept their initial energies from the interval $[-1.5\Phiv;-\Phiv]$, their $r_0$ would lie in
$r_0\in[\frac23 \rvir;\rvir]$.

A question appears: how strong can the energy evolution of the system during its formation be,
i.e., can the relaxation be arbitrarily strong, or the ratio between the final $\ep_f$ and the
initial $\ep_i$ energies of the particles is somehow limited for the majority of DMPs? In practice,
two scenarios are possible:
\begin{enumerate}
 \item \label{19item1} \emph{Strong energy relaxation.} The particle energies are completely redistributed, final particle energies cower all the
 potential well $[\phi(0);-\Phiv]$. Then $r_0$  of the particles cover all the possible interval $[0,\rvir]$
 more or less evenly.
\item \emph{Moderate energy relaxation} \citep{15}. The final total specific energy $\ep_f$ of most
of the particles differs from the initial ones $\ep_i$ no more than by a factor $k$
\begin{equation}
\dfrac{\ep_f}{\ep_i}\le k \label{19d1}
\end{equation}
and $k\ll\phi(0)/\Phiv$. There can be particles that have changed their energy stronger, but their
fraction is respectively small.\label{19item2}
\end{enumerate}

The relaxation is moderate if $k\le \cvir/4$ \citep{16}. This condition is very strict for galaxy
clusters, but quite soft for low-mass galaxies with $c_{vir}>20$ (though the real density profile
may significantly differ from the NFW one, we used here the NFW halo concentration $\cvir$ because
of its popularity and in view of the fact that characteristic values of $\cvir$ for various types
of astronomical objects are well known).

The moderate relaxation leads to a very important consequence. Since initial particle energies were
close to $-\Phiv$ and changed no more than $k$ times during the halo formation, the halo contains
only a small fraction of particles with the energies lower than $-k\Phiv$ (it is important that
$|k\Phiv|\ll |\phi(0)|$), which roughly corresponds to $r_0\simeq \rvir/k$. Particles with more
compact orbits are almost absent in the halo. It means that the density profile in the center of
the halo, inside $\rvir/k$, is mainly formed by the particles that arrive there from the outside.
It leads to a very peculiar structure of the clump, differing drastically from the clumpy one.

First of all, it is easy to show that a cusp may form only in scenario \ref{19item1}, i.e., if the
relaxation is strong \citep{15, 16}. If the relaxation is moderate, the density profile obligatory
has a core. Nevertheless, the profile still can be very concentrated: the ratio of the core radius
to the virial one can be arbitrarily small. The velocity distribution in the case of moderate
relaxation is isotropic only in the very center of the halo \citep{14} and gains some anisotropy
with radius.

What scenario, \ref{19item1} or \ref{19item2}, is realized in nature? As of now, the answer to this
question is not known with certainty. The mass hierarchy of the real Universe is so vast (from rich
galaxy clusters of mass $\sim 10^{15} M_\odot$ down to clumps of mass $10^{-6} M_\odot$ or even
lower) that it can be covered in no simulation, and a far interpolation is necessary. It makes
discussible some simulation results concerning the clumps. For instance, if we interpolate the
power-law grows of halo concentration $c_{vir}$ with its mass decreasing \citep{navarro2010}, we
obtain extremely large concentrations for clumps. This result is discussable \citep{diemand2013}.
Recent N-body simulations \citep{sanchez2014} suggest that the power-law grows of $c_{vir}$ breaks
for small-mass halos. Moreover, the simulations may suffer from underestimated numerical effects in
halo centers \citep{13}. N-body simulations favor strong energy relaxation leading to the cuspy
profile. Their results nearly approximate galaxy clusters. However, if we consider smaller halos,
like spiral galaxies or dwarf spheroidals, the model of moderate relaxation describes observations
much better \citep{15,16}. In particular, observations suggest cored profiles \citep{deblok2001,
bosma2002, marchesini2002, gentile2007, mamon2011, oh2011}. Therefore, we may expect that formation
of low-massive clumps also follows the moderate relaxation scenario.

The moderate energy evolution can be characterized by the above-mentioned relaxation parameter $k$.
As we will see in the {\it Comparison of the predictions of core and cusp models} section, the
larger $k$ is, the stronger the relaxation is, and the more stable the clumps are. The clumps are
the most fragile if $k=1$, i.e., when the clumps are formed with a negligible relaxation, and
therefore their particles have $r_0\sim \rcl$. It is reasonable to expect that the moderate
relaxation is much stronger ($k\simeq\cvir/4$) \citep{16}. However, in this paper we will mainly
consider the case $k=1$ for simplicity, as it gives the most pessimistic estimation of clump
survival. Thus, though the choice between scenarios \ref{19item1} and \ref{19item2} is now unclear,
the model we consider in this paper may be used to estimate the minimal value of the boost factor.

Moreover, if the assumption of the moderate energy evolution is true, it suggests that the majority
of the particles still concentrate near $\ep=-\Phiv$: the energy exchange of the particles during
the relaxation is more or less a stochastic process, while the total energy of the system should
conserve. Thus the narrowness of the initial energy distribution, the relative smallness of the
energy evolution, and the fact that the final potential well is much deeper, than the initial one,
together result in crowding of the particle apocentre distances $r_0$ near $\rvir$, i.e., $\langle
r_0\rangle\sim\rcl$.

\begin{figure}[tb]
\centerline{\includegraphics[width=0.5\textwidth]{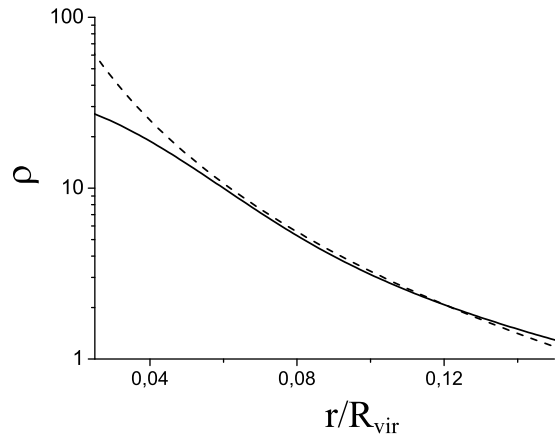}} \caption{The profile of a clump
formed in the case if its energy relaxation was moderate (solid line). The NFW profile with the
same concentration parameter $c_{vir}=25$ is reproduced for comparison (dashed line). The density
is represented in arbitrary units.}\label{19fig1}
\end{figure}

\section{Typical clump parameters}
If the dark matter particle is a WIMP of mass exceeding $100$~{GeV}, the minimal clump mass is
roughly equal to $10^{-6} M_\odot$, which corresponds to $\rvir\sim 0.05$~{pc}. The average density
$\bar \rho$ of the clump is defined from $\bar \rho\cdot \frac43 \pi\rvir^3=\mvir$. We will use the
following designations: $m_6=\mvir/10^{-6} M_\odot$, $r_6=\rvir/0.05$~{pc},
$\bar\rho_6=1.9\cdot10^{-3}$~$M_\odot/\text{pc}^3$. For a clump with mass $\mvir= 10^{-6}M_\odot$
we have $m_6=r_6=\bar\rho_6=1$.

As we could see, almost all the particles in our model have $r_0\sim \rvir$, and the dynamical time
of the clump can be estimated as $t_f\sim\sqrt{\dfrac{\rvir^3}{G\mvir}}\propto(G\bar \rho)^{-1/2}$.
If $\mvir= 10^{-6}M_\odot$, $t_f\sim 10^8$~ years. In the standard $\Lambda$CDM cosmology $\bar
\rho$ depends on the clump mass only logarithmically \citep{gorbrub2, 16}, and as a result, the
clumps of all masses have comparable dynamical times. The time is significantly larger than typical
times of any clump destruction mechanisms. Therefore, the clump disintegration can be considered in
the instant tide approximation.

\section{Clump destruction: the case of the most fragile clumps $k=1$} There are several mechanisms
of clump destruction in our Galaxy. We will consider only the most important ones. As we showed at
the close of section {\it Energy structure of clumps}, the case $k=1$ corresponds to $r_0\sim \rcl$
and gives the most pessimistic estimation of clump survival. We will actively use this fact in the
current section.

\subsection{Destruction on halo stars}
The clumps can be destroyed by collisions with stars. Let us denote the average star mass, star
number density and star velocity dispersion by $m_*$, $n_*$, and $\sigma_*$, respectively. The
dispersion of clump velocities is $\sigma_{cl}$. We may introduce $\sigma =\sqrt{\sigma^2_* +
\sigma^2_{cl}}$. The characteristic time of clump destruction by collisions with stars is
\citep[eqn. 8.55]{bt}
\begin{equation}
\label{19a1}
 t_d\simeq \frac{17}{200}\frac{\sigma \mcl \rcl^2}{G m^2_* n_* r_0^3},
\end{equation}
Since $r_0\sim \rcl$, we obtain:
\begin{equation}
\label{19a2}
 t_d\simeq \frac{17}{200}\frac{\sigma \mcl}{G m^2_* n_* \rcl},
\end{equation}
The main fraction of time the clumps spend in the Milky Way halo. The star concentration there can
be estimated \citep{berezinsky2008} as $n_*=(\rho_h/m_*) (r_\odot/r)^3$, where $r_\odot=8.5$~{kpc},
$\rho_h=1.4\cdot 10^{-5}M_\odot/\text{pc}^3$, $m_*\simeq 0.4 M_\odot$. We should integrate $n_*$
over the clump trajectory, from the minimum $r_{min}$ to the maximum $r_{max}$ radius of its orbit.
However, $n_*$ rapidly drops with radius, and the region near $r_{min}$ makes the main contribution
into the clump destruction. Assuming $\sigma=200$~{km/s}, we obtain from (\ref{19a2}):
\begin{equation}
\label{19a3}
 t_d\simeq (r/r_\odot)^3 m^{2/3}_6 \cdot 8\cdot 10^{10} \text{years}.
\end{equation}
As we can see, $t_d$ exceeds the age of the Universe even for low-massive objects. Only very small
clumps $\mvir< 10^{-7} M_\odot$ can be destroyed by the halo star collisions.

\subsection{Destruction on the Milky Way disc}
However, the clumps can be disrupted by the tidal perturbations of the galactic disc as a whole. A
single passing of a clump through the disc changes its energy on
\begin{equation}
\label{19a4}
 \delta E=\frac{16\pi^2 G^2 o^2_d r_0^2}{3 v_\perp^2},
\end{equation}
where $o_d$ is the surface density of the Milky Way disc, $v_\perp$ is the perpendicular to the
disc component of the clump velocity. We assume \citep{berezinsky2008}
\begin{equation}
\label{19a5}
 o_d(r)=\frac{M_d}{2\pi r^2_d}e^{-r/r_d},
\end{equation}
with $M_d=8\cdot 10^{10}M_\odot$, $r_d=4.5$~{kpc}. Dividing the total clump energy
$|E|\sim\dfrac{G\mcl}{\rcl}$ by (\ref{19a4}), we obtain the number of passing $N_d$ needed to
destroy the clump by the disc
\begin{equation}
\label{19a6}
 N_d=\frac{3 r^4_d v_\perp^2}{8 G M^2_d} \:\frac{\mcl}{\rcl^3}\:e^{2r/r_d}.
\end{equation}
Substituting\footnote{Strictly speaking, $v_\perp$ depends on radius. We may roughly estimate it in
the following way: since the DM halo is more or less spherically symmetric, $v^2_\perp\sim v^2/3$,
and we may assume that $v$ is approximately equal to the circular orbital speed $v_{orb}$.
Generally speaking, $v_{orb}$ also depends on radius (to be more precise, on the mass inside the
radius, which is not well known for $r\sim 50$~{kpc}). However, if the density profile can be
approximated as $\rho\propto r^{-2}$, $v_{orb}={\it const}$. Since our calculation is estimative,
we assume for simplicity that $v_\perp$ is constant and $v^2_\perp=v^2_{orb}(r_\odot)/3\simeq
220^2/3\simeq 10^4$~{(km/s)$^2$}.} here $v_\perp\sim 100$~{km/s}, we obtain $N_d\simeq 4.5\cdot
10^{-4} \bar\rho_6 e^{2r/r_d}$. As we can see, $N_d$ depends only on the average density of the
clump. Since $\bar \rho$ depends on the clump mass only logarithmically in the standard
$\Lambda$CDM model \citep{gorbrub2, 16}, the clump destruction by the disc depends almost not at
all on the clump mass.

The motion of a clump in the Galaxy can be characterized by the minimum $r_{min}$ and the maximum
$r_{max}$ radii of its orbit. Equation (\ref{19a6}) shows that a single passing through the disk is
enough to disintegrate a clump if $r_{min}<17$~{kpc}. So high efficiency of the mechanism under
consideration in the central area of the Milky Way justifies our neglect of other clump destruction
factors, like the galactic bulge, gas nebulas etc. In the regions where they are effective, the
clumps are disrupted anyway.

On the other hand, the disc surface density rapidly drops with radius, and so does the efficiency
of clump destruction. A clump passes through the disc twice in each revolution. Let us consider a
clump with a circular orbit of radius $25$~{kpc}. The orbital period of the Solar System about
Galactic Center is $T_{MW}\sim 230$~{Myr.} If we accept that the Milky Way density profile is
$\rho\propto r^{-2}$ (which is a very rough approximation), $T\propto r$ \citep{11}, i.e., the
orbital period of the clump is $T\simeq 230\text{Myr}\cdot 25\text{kpc}/8\text{kpc}\simeq
700$~{Myr.} The clump passed through the disc $\sim 30$ times in the lifetime of the Galaxy.
According to Eqn.~\ref{19a6}, the clump should be destroyed in $\sim 30$ passages. Thus
$r=25$~{kpc} is approximately the border of the region where the clumps were disrupted by the disc
in the lifetime of the Galaxy. Clumps with larger $r$ also can be destroyed, if they have an
elongate orbit and pass through the disc close to the center. However, the fraction of these clumps
falls approximately as $r^{-2}$ and becomes negligible at $r\sim 50$~{kpc}.

\subsection{Destruction by the gravitational field of the Galaxy}
If a clump has a non-circular orbit, it periodically approaches the galaxy center and expresses
tidal shocks. A single shock increases the clump energy on \citep[eqn. 8.43]{bt}
\begin{equation}
\label{19a7}
 \delta E=\frac{4 G^2 M^2_p(r_{min})}{3 v^2 r_{min}^4}\langle r_0^2\rangle,
\end{equation}
where $v$ is the clump velocity at $r_{min}$, and $M_p(r_{min})$ is the total mass of the perturber
(in our case, of the Milky Way) inside $r_{min}$, and $\langle r^2\rangle$ is the mass-weighted
mean-square radius of the clump. If the clump density profile is $\rho\propto r^{-2}$ and
$r_0=\rcl$, $\langle r^2\rangle=r_0^2/3=\rcl^2/3$. Though a real clumps apparently have other
profiles, we may use approximation $\langle r^2\rangle=\rcl^2/3$ in our estimative calculations.
Dividing the clump energy $E\simeq G\mcl/\rcl$ by $\delta E$, we obtain the number of rotations
necessary to destroy the clump
\begin{equation}
\label{19a8}
 N_d= \frac94 \left(\frac{v^2 r_{min}}{G M_p(r_{min})}\right) \left(\frac{\mcl}{M_p(r_{min})}\right)
 \left(\frac{r_{min}}{\rcl}\right)^3.
\end{equation}
If the orbit has a significant eccentricity, we may estimate $v^2\simeq G M_p(r_{min})/r_{min}$.
\begin{equation}
\label{19a9}
 N_d(r)\simeq 2 \frac{\bar \rho_{cl}}{\bar\rho_p(r)},
\end{equation}
where $\bar\rho_p(r)$ is the average density if the perturber inside radius $r$. Thus the clump can
be effectively destroyed if it passes through the areas where its average density is comparable
with the density of the perturber. The destruction process again depends only on $\bar \rho_{cl}$
and therefore is almost insensitive to the clump mass. According to equation (\ref{19a9}), a clump
should have orbit radius $r\sim 18$~{kpc} do be destroyed in one turn about the Galaxy center and
$r\sim 50$~{kpc} to be destroyed in the lifetime of the Galaxy, for the conventional Milky Way
models.

We should underline that the mechanism efficiency strongly depends on the orbit eccentricity: a
clump on a circular or near-circular orbit expresses no shocks. However, there is a clump
destruction mechanism that is not sensitive at all to the shape of the clump orbit. Indeed, let us
consider a particles belonging to clump that rotates around its center having the apocenter
distance $r_0$. The tidal force from the Galaxy gravitational field can be estimated as $(G
M_p(r)/r^2)\cdot (r_0/r)$ in the framework of the clump center, where $r$ is the clump distance
from the Galaxy center. The work of the force on the particle orbit is $\sim r_0\cdot(G
M_p(r)/r^2)\cdot (r_0/r)$. If it is comparable with the bind energy of the particle $\sim
G\mcl/r_0$, the particle can be teared out of the clump. We assume $r_0\sim \rcl$, as we did
before. The condition of clump destruction is
\begin{equation}
\label{19a10}
 \frac{M_p(r)}{r^3}>\frac{\mcl}{\rcl^3} \qquad\text{or}\qquad \bar \rho_{cl}< \bar\rho_p(r).
\end{equation}
If we compare equation~(\ref{19a9}) with the last one, we can see, that it is approximately equal
to the condition for a clump to be destroyed in one or two turns around the Galaxy center. Thus the
direct tidal disruption is in general less effective than the tidal shocks. On the other hand,
equation~(\ref{19a10}) guarantees that all the clumps (in the most pessimistic case for their
survival, when $r_0\sim \rcl$) are destroyed inside $r\sim 20$~{kpc}, disregarding their orbits.

To summarize: The clumps in the Milky Way halo are destroyed by the disc and by tidal effects in
the Galaxy gravitational field. The efficiency of these two mechanisms is comparable. Collisions
with halo stars are not effective. All the clumps should be destroyed inside $\sim 20$~{kpc} from
the Galaxy center (we should remind that we consider the most pessimistic scenario). A gray zone
lies between $20$ and $50$~{kpc}, some clumps should be destroyed there, some may survive. No
mechanism effectively disrupts clumps out of $50$~{kpc} from the Milky Way center, even under the
most pessimistic assumptions about the inner clump structure that we consider.

\subsection{Dwarf spheroidals. Segue 1.} Among all astrophysical objects containing dark matter,
dwarf satellites of the Local Group draw special attention. They are very promising for detecting
dark-matter annihilation. Though the dark-matter density in these objects is significantly lower
than that at the Galactic center, they typically contain no sources of cosmic rays, and therefore
almost any high-energy signal detected from the dwarf galaxies could be the signal of dark-matter
annihilation.

Segue 1 draws the most attention among numerous satellites in the Local Group \citep{geringer2015}.
It combines several advantages: it is the closest satellite to Earth, it is located at high
Galactic latitude (and, consequently, superposes on a low gamma-ray background) and expected to
produce the strongest dark-matter signal \citep{strigari2010}, though the last statement is still
discussable. Recent observations can essentially constrain the most probable WIMP dark matter
models (with the cross-section corresponding to the dark-matter abundance $\langle \sigma
v\rangle_a \simeq 3\cdot 10^{-26} \text{cm}^3/\text{s}$ and DMP mass range that has not been
excluded by LHC experiments) only if the dwarf spheroidals are expected to have a significant boost
factor \citep{12}. Therefore the question of the substructures in dwarf spheroidals is extremely
important.

Segue 1 has the highest known mass-to-light ratio of any observed galaxy, which implies that Segue
1 is strongly dominated by dark matter. Its density distribution can be modelled by the Einasto
profile
\begin{equation}
\label{19a11}
 \rho=\rho_e \exp\left[-2 n\left\{\left( \frac{r}{r_e}\right)^\frac{1}{n}
 -1\right\}\right]
\end{equation}
with $\rho_e=0.11$~$M_\odot$\,pc$^{-3}$, $r_e=150$~{pc}, and $n=3.3$ \citep{magic}. Unfortunately,
all the parameters are poorly known and contain huge uncertainties: all the stars ($\simeq 66$)
observed in Segue 1 lie inside $10'$ ($\simeq 67$~pc) from the center, which is two times smaller
than $r_e$. For instance, a completely different set of parameters $\rho_e= 0.094$~$M_\odot$
\,pc$^{-3}$, $r_e=70$~{pc} fits the observations equally well \citep{segue12010}.

Apparently, Segue 1 does not contain a disc. However, the clumps could still be disrupted by stars
or by the tides in the gravitational field of the dwarf. We consider both these mechanisms.

Segue 1 contains approximately a thousand stars \citep{frebel2014} that are situated inside $\sim
67$~{pc} from the center. We may very roughly estimate $n_*\simeq 8\cdot 10^{-4} \text{pc}^{-3}$.
Substituting $m_*=0.4 M_\odot$ and $\sigma=60$~{km/s} (which is approximately the escape velocity
from the center of Segue 1 \citep{12}), we obtain
\begin{equation}
\label{19a12}
 t_d\simeq m^{2/3}_6 \cdot 10^{9} \text{years}.
\end{equation}
As we can see, the smallest clumps are destroyed by the stars. However, clumps of mass $m_6>30$,
i.e., $m>3\cdot 10^{-5} M\odot$, cannot be disrupted by this mechanism.

In contrast, the clump destruction by the gravitational field of Segue 1 is extremely effective. We
can see from eqn.~(\ref{19a11}) that the equality $\bar \rho_{cl}=\bar\rho_p(r)$ occurs at the
distance $\sim 1300$~{pc} from the center of Segue 1 (if we suppose that mass
distribution~(\ref{19a11}) is still valid there).

\section{The case of robust clumps $k=\cvir/4$}
The case of $k=1$ is extreme and not very probable, even if the relaxation is moderate. One may
expect that $k$ depends on $\cvir$ of the halo. The most probable value of $k$ is $\cvir/4$
\citep{16}, which is close to the maximum value of $k$, at which the relaxation may still be named
moderate. Therefore we should consider the case $k=\cvir/4$.

First of all, now $k$ depends on $\cvir$. If $k=\cvir/4$, $\langle r_0\rangle \gtrsim \rcl/k =
4\rvir/\cvir$ \citep{15}, and we should substitute this value to (\ref{19a7}) and (\ref{19a4}).
Recent simulation suggest that the power-law growth of $\cvir$ with the halo mass decreasing
breaks, and the concentration of low-mass objects grows much slower. A sophisticated analysis
\citep{sanchez2014} has shown that even the smallest halos have $\cvir\simeq 60$. We will use this
value as the most plausible.

Equation (\ref{19a6}) for the number of rotations necessary to destroy the clump by the Milky Way
disc transforms into
\begin{equation}
\label{19d2}
 N_d=\frac{3 r^4_d v_\perp^2 \cvir^2}{128 G M^2_d} \:\frac{\mcl}{\rcl^3}\:e^{2r/r_d}.
\end{equation}
A reasoning similar to that used below equation (\ref{19a6}) shows that $r=15$~{kpc} is
approximately the border of the region where the clumps were disrupted by the disc in the lifetime
of the Galaxy in the case of robust cored clumps with $k=\cvir/4$. This value is substantially
smaller than that for the most fragile clumps $k=1$, but larger then the value obtained for the
cuspy clumps $r\sim 8$~{kpc} \citep{berezinsky2008}. Thus the durability of the cored clumps may
vary in a wide range. However, an essential difference between cuspy and cored models remains.

Equation (\ref{19a10}) transforms into
\begin{equation}
\label{19a13} \frac{\cvir^2}{16} \rho_{cl}< \bar\rho_p(r).
\end{equation}
This equation shows that the dwarf satellites like Segue 1 are so dense that it is difficult to
expect that any cored clumps could survive in it. Indeed, if we accept $\cvir\simeq 60$ for clumps,
then the clumps are destroyed in the region where $\bar\rho_p(r)> 225 \rho_{cl}$.
Equation~(\ref{19a11}) shows that all the clumps should be destroyed inside $\sim 2 r_e$, i.e.,
inside $300$~{pc} from the center of Segue 1.

The conclusion of absence of any DM substructures is probably valid for all dwarf satellites of the
Local Group. The satellites, as we observe them now, are the most probably the central and the
densest parts of ancient dwarf galaxies tidally destroyed by the Milky Way and M31. As a result,
their density is very high, and the clumps are easily destroyed when approaching their centers. We
should also take into account that the tidal perturbations destroying the satellites disrupted
their clump structure as well.
\begin{figure}[tb]
\centerline{\includegraphics[width=0.5\textwidth]{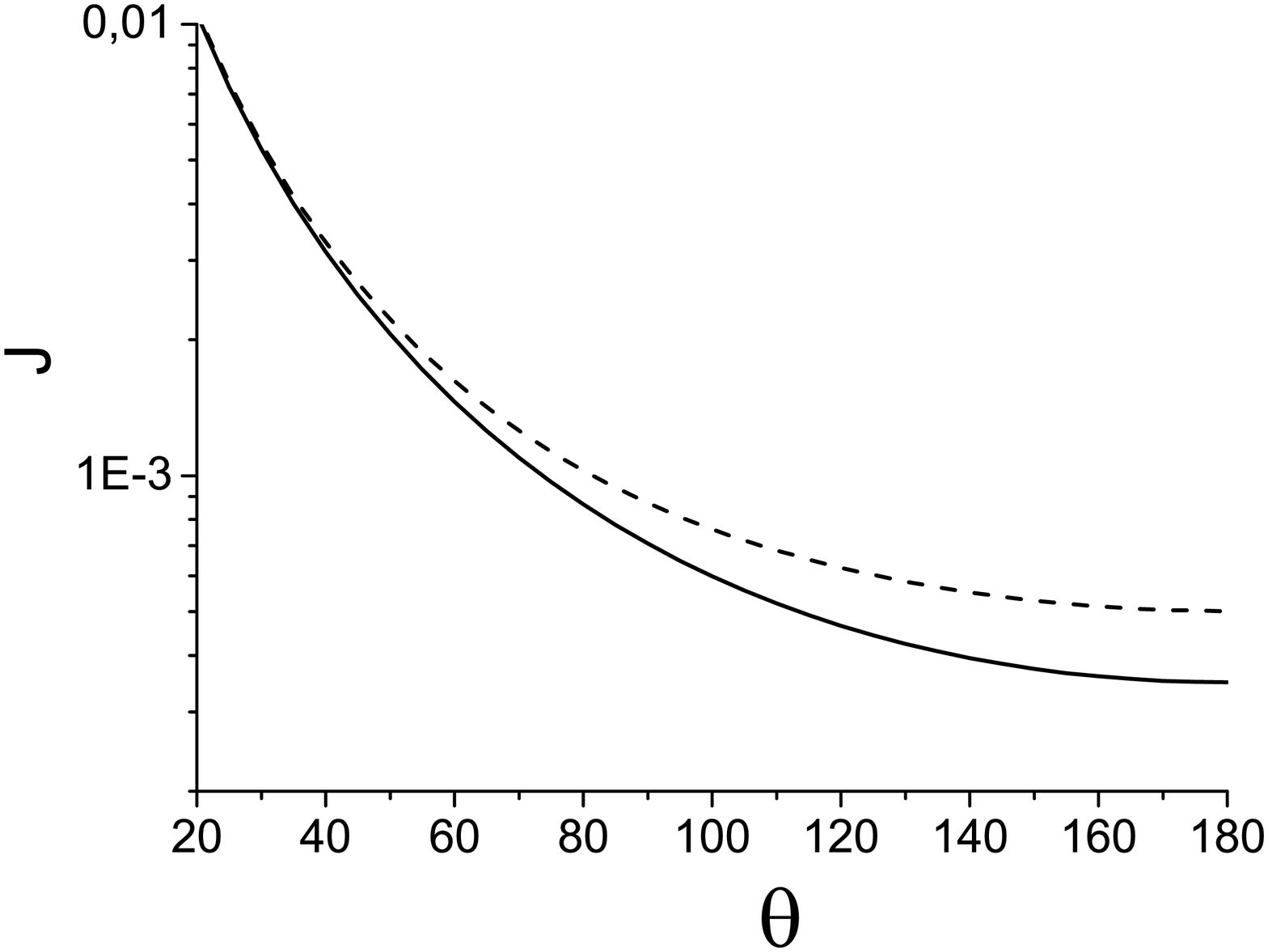}} \caption{The annihilation signal
$J$ from the Milky Way as a function of the angle $\theta$ between the observation line and the
direction towards the Galaxy center. The normalization of $J$ is arbitrary. The solid line
corresponds to the case when the clumpy structure disappears at $r_{destr}=50$~{kpc}, the dashed
line - to $r_{destr}=10$~{kpc}. The boost factor at $50$~{kpc} is set to be equal to
$C_{50}=20$.}\label{19fig3}
\end{figure}

\section{Comparison of the predictions of core and cusp models}
Now we can compare the survival of the cored clumps formed under the assumption of moderate
relaxation (hereafter MR-clumps) with one of the standard model NFW clumps \citep{berezinsky2006,
berezinsky2008}. As we can see, clumps are destroyed much easier in our case.

This result should not surprise. The clumps of the standard model are significantly more stable not
because of the existence of cusp per se: MR-clumps can also be very concentrated and have large
central density (Fig~(\ref{19fig1})). The principle difference is the particle orbits. The
particles forming the cusp in the NFW model are confined in the cusp: they have small
quasi-circular orbits with the apocenter distance $r_0\simeq 0$. According to eqn.~\ref{19a4}, the
energy perturbations are proportional to $r^2_0$. This fact along provides the cusp stability.
Moreover, the particles with $r_0\simeq 0$ have a very short dynamical time, the tidal
perturbations are adiabatic for these particles, and equations like (\ref{19a6}) or (\ref{19a9})
are not valid anymore. The adiabatic invariant conservation makes the cusp extremely resistant to
tidal perturbations. A NFW clump easily looses its outer layers, but the cusps are almost immune.
They can be destroyed only in the bulge of the Milky Way, at the distances less then $\sim 3$~{kpc}
from the Galaxy center \citep{berezinsky2008}, and most likely survive in the dwarf spheroidals.
Since it is cusps that makes the main contribution into the boost factor, it should be significant
in the standard model.

The MR-clumps are built completely differently. They still can have a very high central density,
and their profile can be relatively close to the NFW, except of a small central region, where they
have cores (see Fig~(\ref{19fig1})). However, the majority of the DMPs have the apocenter radius
$r_0$ comparable with the virial radius of the clump. It is true even for the core itself: The bulk
of the particles only 'visit' the halo center, having $r_0 \gg r_c$, where $r_c$ is the core
radius. As we could see, the particles with large apocenter distance $r_0$ are much easier for the
tidal effects to detach. The MR-clumps have a much larger apocenter radius $\langle r_0\rangle$ and
therefore are much easier to be destroyed.

Evidently the largest possible value $\langle r_0\rangle=R_{vir}$, which corresponds to the
relaxation parameter $k=1$. It is exactly the assumption we used in obtaining of the results of
this paper. Therefore we have obtained the low limit on the clump stability. The most probable
value of $k$ is $k\simeq\cvir/4$, where $\cvir$ is the NFW halo concentration \citep{16}. Then
$\langle r_0\rangle\simeq 2 R_{vir}/\cvir$, and the clump survival is somewhat higher. However, it
is still much lower for the cored clumps than for the cuspy ones.

We have shown that the clump survival strongly depends on their internal structure, which is not
quite clear now. Since the annihilation signal rapidly increases with the DM density growth,
compact and dense DM regions attract the main attention of the indirect DM searches. Our
calculations show that the clump structure in these objects can be utterly destroyed under quite
natural cosmological assumptions.

It is important that, even under the most pessimistic suppositions about their structure, the
clumps in the Milky Way should survive at the distances exceeding $50$~{kpc} from the Galaxy
center. It means that the clumpy structure should survive in almost all the dark matter in the
Universe. If we consider the Local Group, $\sim 80\%$ of its DM containment form the halos of the
Milky Way and M31 outside $50$~{kpc} from their centers and probably a common DM envelope of the
system \citep{loeb2008}. Broadly speaking, the main part of the dark matter in the Universe forms
low-density structures, like DM envelopes of galaxy clusters, walls and filaments of the large
scale structures etc. All this dark  matter should have kept entirely its clumpy structure. This
fact is very important for the estimations of the extragalactic background, possibly created by the
DM annihilation.

\section{Phenomenology: predictions of the annihilation signal and uncertainties}
 We should underline that the phenomenological consequence of the DM substructures is not only
the annihilation signal boosting by some factor. The signal dependence on the DM density $\rho$
changes. If the dark matter is homogeneous, the annihilation signal is proportional to $\rho^2$. If
the annihilation in clumps dominates, the signal is proportional to the number density of the
clumps, i.e., to $\rho$. Thus, if the clumps exist in the present-day DM structures, the DM
annihilation should be boosted, and the boost factor should be roughly proportional to $\rho^{-1}$.

Let us discuss the uncertainties and possible observable consequences of the model. As we have
shown, the clumps survive outside of $50$~{kpc} from the galactic center even under the most
pessimistic supposition about their structure. On the other hand, \citep{berezinsky2008} showed
that even rather robust NFW clumps are strongly destroyed closer than $8$~{kpc} from the center. We
may conclude that $20-50$~{kpc} is quite a reliable estimation of the maximum distance from the
Milky Way center where the cored clumps can still be destroyed. The boost factor estimations are
much more ambiguous.

\subsection{A sketch of the standard theory of the boost factor}

If the comoving number density of halos in the mass range of $[M; M+dM]$ at the redshift $z$ is
$dp(M,z)/dM$, the boost factor of the system $C$ can be  represented as a multiplication (see
details in \citep{ahn2005}):
\begin{equation}
  C =\Delta \cdot F_{coll}(z) \cdot \left[B_{halo}\right]. \label{19b2}
\end{equation}
Here $\Delta$ is the mean halo overdensity defined by the cosmological model, of a halo in units of
the cosmic mean matter density, $F_{coll}(z)=\int \frac{dp}{dM} M dM/\langle \rho \rangle$ is the
mass fraction collapsed into cosmological halos at $z$ ($\langle \rho \rangle$ is the average
density of the Universe). $B_{halo}$ is a single halo clumping factor $B_{halo}= \int \left(
\frac{\rho}{\langle \rho \rangle_{halo}} \right)^{2} d^{3}r  / \int d^{3}r$, where $\langle \rho
\rangle_{halo}$ is the halo average density, and $\left[B_{halo}\right]\equiv \int B_{halo}
\frac{dp}{dM} M dM / \int \frac{dp}{dM}M dM$.

As we can see, the boosting is determined by three functions: the halo mass function $p(M,z)$, the
boost for a single halo $B_{halo}$ as a function of the halo concentration $c_{vir}$, and the
concentration dependence $c_{vir}(M,z)$ of halo mass and $z$.
\begin{equation}
\label{19b3}
B_{halo}=\frac{c_{vir}^{3}(1-1/(1+c_{vir})^3)}{9(\ln(1+c_{vir})-c_{vir}/(1+c_{vir}))^2}
\end{equation}
for the case of NFW halos.

\begin{figure}[tb]
\includegraphics[width=0.5\textwidth]{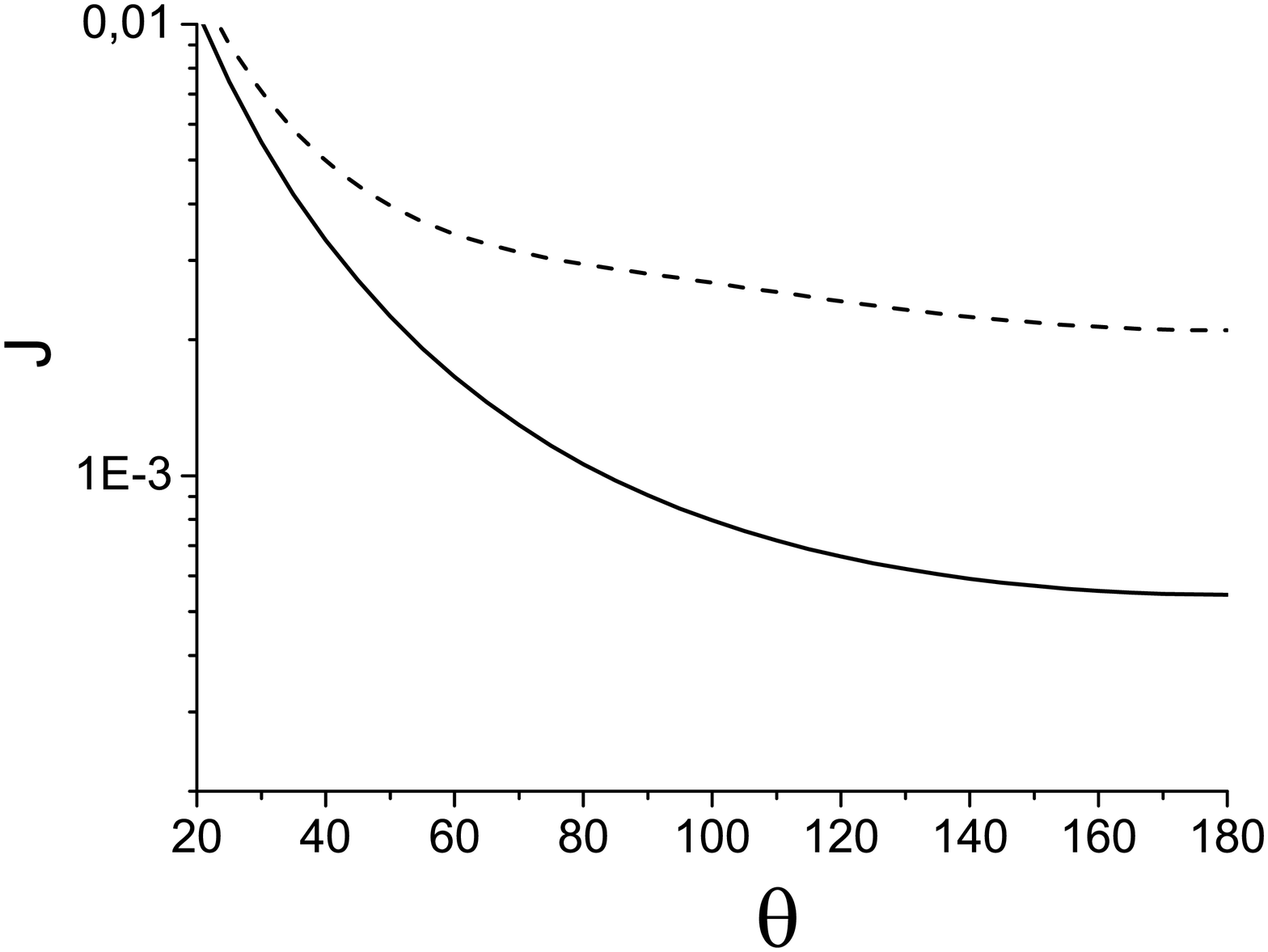}
\caption{The annihilation signal $J$ from the Milky Way as a function of the angle $\theta$ between
the observation line and the direction towards the Galaxy center. The normalization of $J$ is
arbitrary. The solid line corresponds to the case when the clumpy structure disappears at
$r_{destr}=50$~{kpc}, the dashed line - to $r_{destr}=10$~{kpc}. The boost factor at $50$~{kpc} is
set to be equal to $C_{50}=100$.}\label{19fig4}
\end{figure}

\subsection{The boosting in the case of cored clumps}
What changes in the picture, if we consider the above-stated model instead of the traditional NFW
one? It is surprising that the mass function should be almost the same for the cuspy and cored
clumps. Indeed, two main factors defining $p(M,z)$ are the spectrum of the initial cosmological
perturbations and the clump destruction during the hierarchical structure formation. The first
factor has nothing to do with the profiles of formed clumps. The second process (in the framework
of the standard $\Lambda$CDM cosmology) is also not sensitive to the clump density profile. The
objects of different masses collapse almost simultaneously in $\Lambda$CDM, and, as we have already
mentioned, the hierarchical destruction time is shorter than the virilization time. Therefore, the
approach used by \citep{berezinsky2006} and considering non-virialized clumps is exactly valid for
in our case and leads to the same mass function.

The only mechanism that significantly changes $dp(M,z)/dM$ and essentially depends on the clump
density profile is the astrophysical clump destruction. Let us consider the Local Group. As we
could see, the clumps situated farther than $50$~{kpc} from the Milky Way center should survive
under the most pessimistic assumptions about their structure. However, $\sim 70\%$ of the Milky Way
dark matter lies outside of this radius \citep{klypin2013}, and we may suppose that the same
situation takes place in the Andromeda galaxy. Moreover, along with the DM haloes of large and
dwarf galaxies, the Local Group may contain an approximately equal quantity of dark matter that is
not bound in the galaxies and presumably forms a large envelope \citep{bt, loeb2008, 14}. Thus,
$\sim 85\%$ of the Local Group utterly kept its clumpy structure, and if we consider a large area
of dark matter, significantly exceeding the maximum halo size, the modification of the averaged
$dp(M,z)/dM$ is not very significant.

Of course, $B_{halo}$  depends on the clump density profile. However, even in the case of the NFW
clump with infinite central density, the annihilation signal ($\propto\rho^2\cdot 4\pi r^2\propto
{\it const}$) has no peculiarity in the center. Since the clear discrepancy between cores and cusps
occurs only at $r\ll r_s$, $B_{halo}$ for these cases may differ only on a factor $\lesssim 2$ (see
{\it minimal} model in \citep{12} as an illustration). Thus the difference in value of $B_{halo}$
between cored and cuspy clumps is much less then the uncertainties of $c_{vir}(M,z)$.

\begin{figure}[tb]
\centerline{\includegraphics[width=0.5\textwidth]{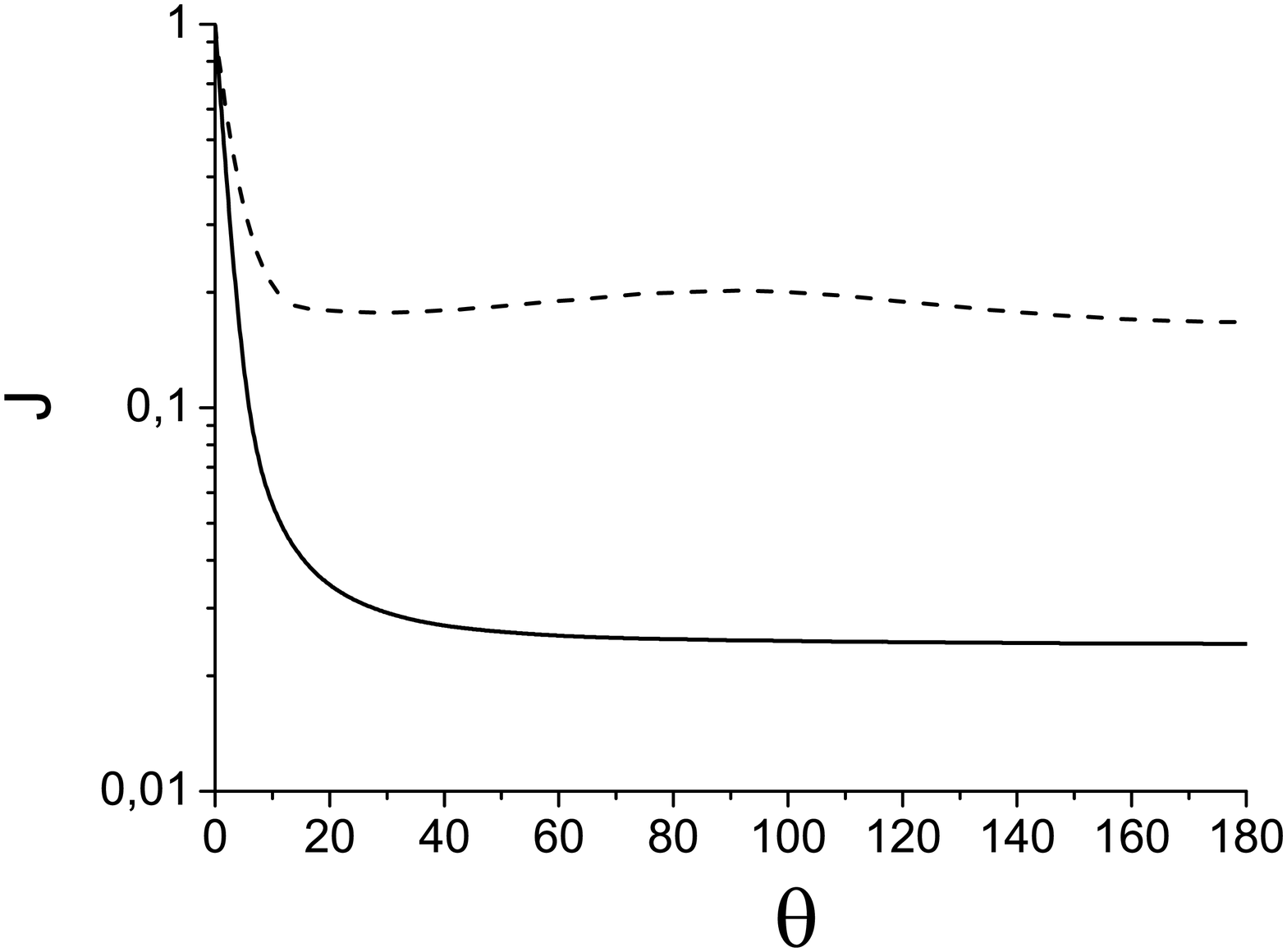}} \caption{The annihilation
signal $J$ from the Milky Way as a function of the angle $\theta$ between the observation line and
the direction towards the Galaxy center. The normalization of $J$ is arbitrary. The solid line
corresponds to the case when the clumpy structure disappears at $r_{destr}=50$~{kpc}, the dashed
line - to $r_{destr}=10$~{kpc}. The boost factor at $50$~{kpc} is set to be equal to
$C_{50}=10^4$.}\label{19fig5}
\end{figure}

Indeed, the uncertainty of $c_{vir}(M,z)$ (especially the dependence on $M$) is the weakest point
of any DM indirect search theory. Since $dp(M,z)/dM$ rapidly drops with $M$, the preponderant
contribution into $\left[B_{halo}\right]$ is given by the smallest clumps, and we need to know
their concentration. Typically $c_{vir}(M,z)$ is defined from the results of N-body simulations
that suggest a power-law dependence \citep{ahn2005}. However, the ratio of masses of the smallest
halos with properly resolved density profiles to the largest ones hardly exceeds $3-4$ orders of
magnitude even in recent large simulations containing $\sim 10^9$ test bodies. The mass hierarchy
of the real Universe covers at least $21$ orders of magnitude. Therefore, the estimations of
$c_{vir}$ for the smallest clumps are based on a far interpolation. It is questionable whether the
power-law behavior of $c_{vir}(M,z)$ holds within such a vast range of masses \citep{diemand2013}.
Recent N-body simulations suggest that $c_{vir}<70$ even for the smallest clumps
\citep{sanchez2014}. Moreover, $B_{halo}\propto c^3_{vir}$ if $c_{vir}\gg 1$, i.e., $B_{halo}$ is
sensitive even to small changes of $c_{vir}$. All these facts make the predictions of
$\left[B_{halo}\right]$ very dependent on the suppositions of the $c_{vir}(M,z)$ behavior. Now we
will discuss several possible scenarios by the example of the Milky Way.

Thus, if we consider a large area of dark matter, significantly exceeding the maximum halo size,
the difference between cuspy and cored scenarios can be surprisingly small, especially if we take
into account the large unavoidable uncertainties. For instance, if we could measure the
extragalactic background produced by the DM annihilation, we would be unable to choose between the
scenarios because of our ignorance about $c_{vir}(M,z)$ of the small clumps. However, if we
consider a separate halo, the predictions of the models can be drastically different.

In order to illustrate this, we consider the Milky Way halo, as it is a popular object for the
indirect DM search. For simplicity, we assume that the clumpy structure appears stepwise at some
radius $r_{destr}$: there is no substructure inside this radius, and the substructures utterly
survive outside of $r_{destr}$. In order to take into account the uncertainties of $c_{vir}$
determination, we presume the total boost factor $C_{50}$ at $r=50$~{kpc} from the Galaxy center.
The absolute value of the signal depends on the particle nature of the DM, and we choose the
normalization arbitrarily (but the same for all the cases). The figures (\ref{19fig3}-\ref{19fig5})
show the annihilation signal $J$ from the Milky Way as a function of the angle $\theta$ between the
observation line and the direction towards the Galaxy center. The solid line corresponds to
$r_{destr}=50$~{kpc}, which is roughly consistent with the scenario of the least stable cored
clumps, the dashed line --- to $r_{destr}=10$~{kpc}, which approximately models the NFW clumps
\citep{berezinsky2008}. Fig.~\ref{19fig3}, \ref{19fig4}, \ref{19fig5} correspond to $C_{50}=20$,
$100$, $10^4$, respectively.

As we can see, the difference between the scenarios is not significant, if the boosting is not very
strong ($C_{50}=20$, Fig.~\ref{19fig3}). The dashed line on Fig.~\ref{19fig4} approximately
coincides with Fig.~5 in \citep{berezinsky2008}: the model with $r_{destr}=10$~{kpc} and
$C_{50}=100$ indeed corresponds to that considered in this paper and assuming the NFW clumps and
holding the power-law behavior of $c_{vir}(M)$ obtained in N-body simulations. The cases of
$C_{50}=20$ and $C_{50}=10^4$ describe slower and faster grows of $c_{vir}(M)$, respectively. One
can see that the difference between the cuspy and cored clumps becomes quite significant if
$C_{50}\ge 100$: the external regions of the halo generate a bright background, and it is
significantly dimer if the clumps are easy to be destroyed. The background significantly modifies
the angular profile of the signal. However, even if the boost factor is as large as $C_{50}=10^4$,
the background produced by the outer clumpy regions of the halo is not bright enough to overshadow
the Galaxy center.

\acknowledgments Financial support by Bundesministerium f\"ur Bildung und Forschung through
DESY-PT, grant 05A11IPA, is gratefully acknowledged. BMBF assumes no responsibility for the
contents of this publication. We acknowledge support by the Helmholtz Alliance for Astroparticle
Physics HAP funded by the Initiative and Networking Fund of the Helmholtz Association. The work is
supported by the CONICYT Anillo project ACT-1122 and the Center of Excellence in Astrophysics and
Associated Technologies (PFB06).


\begin{thebibliography}{10}

\bibitem{gorbrub2}
D.~S. {Gorbunov} and V.~A. {Rubakov}, {\em Introduction to the Early Universe
  theory. Volume 2: Cosmological perturbations.}
\newblock LKI publishing house, Moscow, 2010.

\bibitem{lowmassclump}
K.~P. {Zybin}, M.~I. {Vysotsky}, and A.~V. {Gurevich}, {\it {The fluctuation
  spectrum cut-off in a neutralino dark matter scenario.}},  {\em Physics
  Letters A} {\bf 260} (Sept., 1999) 262--268.

\bibitem{highmassclump}
S.~{Hofmann}, D.~J. {Schwarz}, and H.~{St{\"o}cker}, {\it {Damping scales of
  neutralino cold dark matter}},  {\em \prd} {\bf 64} (Oct., 2001) 083507,
  [\href{http://xxx.lanl.gov/abs/astro-ph/0104173}{{astro-ph/0104173}}].

\bibitem{diemand2013}
D.~{Anderhalden} and J.~{Diemand}, {\it {Density profiles of CDM microhalos and
  their implications for annihilation boost factors}},  {\em \jcap} {\bf 4}
  (Apr., 2013) 9, [\href{http://xxx.lanl.gov/abs/1302.0003}{{arXiv:1302.0003}}].

\bibitem{berezinsky2006}
V.~{Berezinsky}, V.~{Dokuchaev}, and Y.~{Eroshenko}, {\it {Destruction of
  small-scale dark matter clumps in the hierarchical structures and galaxies}},
   {\em \prd} {\bf 73} (Mar., 2006) 063504,
  [\href{http://xxx.lanl.gov/abs/astro-ph/0511494}{{astro-ph/0511494}}].

\bibitem{diemand2005}
J.~{Diemand}, B.~{Moore}, and J.~{Stadel}, {\it {Earth-mass dark-matter haloes
  as the first structures in the early Universe}},  {\em \nat} {\bf 433} (Jan.,
  2005) 389--391, [\href{http://xxx.lanl.gov/abs/astro-ph/0501589}{{
  astro-ph/0501589}}].

\bibitem{berezinsky2008}
V.~{Berezinsky}, V.~{Dokuchaev}, and Y.~{Eroshenko}, {\it {Remnants of dark
  matter clumps}},  {\em \prd} {\bf 77} (Apr., 2008) 083519,
  [\href{http://xxx.lanl.gov/abs/0712.3499}{{arXiv:0712.3499}}].

\bibitem{15}
A.~N. {Baushev}, {\it {Galaxy Halo Formation in the Absence of Violent
  Relaxation and a Universal Density Profile of the Halo Center}},  {\em \apj}
  {\bf 786} (May, 2014) 65, [\href{http://xxx.lanl.gov/abs/1205.4302}{{
  arXiv:1205.4302}}].

\bibitem{suchkov}
L.~S. {Marochnik} and A.~A. {Suchkov}, {\em {The Galaxy}}.
\newblock 1984.

\bibitem{16}
A.~N. {Baushev}, {\it {Relaxation of dark matter halos: how to match
  observational data?}},  {\em \aap} {\bf 569} (Sept., 2014) A114,
  [\href{http://xxx.lanl.gov/abs/1309.5162}{{arXiv:1309.5162}}].

\bibitem{14}
A.~N. {Baushev}, {\it {Extragalactic Dark Matter and Direct Detection
  Experiments}},  {\em \apj} {\bf 771} (July, 2013) 117,
  [\href{http://xxx.lanl.gov/abs/1208.0392}{{arXiv:1208.0392}}].

\bibitem{navarro2010}
J.~F. {Navarro}, A.~{Ludlow}, V.~{Springel}, J.~{Wang}, M.~{Vogelsberger},
  S.~D.~M. {White}, A.~{Jenkins}, C.~S. {Frenk}, and A.~{Helmi}, {\it {The
  diversity and similarity of simulated cold dark matter haloes}},  {\em
  \mnras} {\bf 402} (Feb., 2010) 21--34,
  [\href{http://xxx.lanl.gov/abs/0810.1522}{{arXiv:0810.1522}}].

\bibitem{sanchez2014}
M.~A. Sánchez-Conde and F.~Prada, {\it The flattening of the
  concentration–mass relation towards low halo masses and its implications
  for the annihilation signal boost},  {\em Monthly Notices of the Royal
  Astronomical Society} {\bf 442} (2014), no.~3 2271--2277,
  [\href{http://xxx.lanl.gov/abs/http://mnras.oxfordjournals.org/content/442/3/2271.full.pdf+html}{{
  http://mnras.oxfordjournals.org/content/442/3/2271.full.pdf+html}}].

\bibitem{13}
A.~N. {Baushev}, {\it {The real and apparent convergence of N-body simulations
  of the dark matter structures: Is the Navarro-Frenk-White profile real?}},
  {\em Astroparticle Physics} {\bf 62} (Mar., 2015) 47--53,
  [\href{http://xxx.lanl.gov/abs/1312.0314}{{arXiv:1312.0314}}].

\bibitem{deblok2001}
W.~J.~G. {de Blok}, S.~S. {McGaugh}, and V.~C. {Rubin}, {\it {High-Resolution
  Rotation Curves of Low Surface Brightness Galaxies. II. Mass Models}},  {\em
  \aj} {\bf 122} (Nov., 2001) 2396--2427.

\bibitem{bosma2002}
W.~J.~G. {de Blok} and A.~{Bosma}, {\it {High-resolution rotation curves of low
  surface brightness galaxies}},  {\em \aap} {\bf 385} (Apr., 2002) 816--846,
  [\href{http://xxx.lanl.gov/abs/astro-ph/}{{astro-ph/}}].

\bibitem{marchesini2002}
D.~{Marchesini}, E.~{D'Onghia}, G.~{Chincarini}, C.~{Firmani}, P.~{Conconi},
  E.~{Molinari}, and A.~{Zacchei}, {\it {H{$\alpha$} Rotation Curves: The Soft
  Core Question}},  {\em \apj} {\bf 575} (Aug., 2002) 801--813,
  [\href{http://xxx.lanl.gov/abs/astro-ph/}{{astro-ph/}}].

\bibitem{gentile2007}
G.~{Gentile}, P.~{Salucci}, U.~{Klein}, and G.~L. {Granato}, {\it {NGC 3741:
  the dark halo profile from the most extended rotation curve}},  {\em \mnras}
  {\bf 375} (Feb., 2007) 199--212,
  [\href{http://xxx.lanl.gov/abs/astro-ph/}{{astro-ph/}}].

\bibitem{mamon2011}
L.~{Chemin}, W.~J.~G. {de Blok}, and G.~A. {Mamon}, {\it {Improved Modeling of
  the Mass Distribution of Disk Galaxies by the Einasto Halo Model}},  {\em
  \aj} {\bf 142} (Oct., 2011) 109,
  [\href{http://xxx.lanl.gov/abs/1109.4247}{{arXiv:1109.4247}}].

\bibitem{oh2011}
S.-H. {Oh}, W.~J.~G. {de Blok}, E.~{Brinks}, F.~{Walter}, and R.~C.
  {Kennicutt}, Jr., {\it {Dark and Luminous Matter in THINGS Dwarf Galaxies}},
  {\em \aj} {\bf 141} (June, 2011) 193,
  [\href{http://xxx.lanl.gov/abs/1011.0899}{{arXiv:1011.0899}}].

\bibitem{bt}
J.~{Binney} and S.~{Tremaine}, {\em {Galactic Dynamics: Second Edition}}.
\newblock Princeton University Press, 2008.

\bibitem{11}
A.~N. {Baushev}, {\it {Principal properties of the velocity distribution of
  dark matter particles on the outskirts of the Solar system}},  {\em \mnras}
  {\bf 417} (Oct., 2011) L83--L87,
  [\href{http://xxx.lanl.gov/abs/1103.3828}{{arXiv:1103.3828}}].

\bibitem{geringer2015}
A.~{Geringer-Sameth}, S.~M. {Koushiappas}, and M.~G. {Walker}, {\it
  {Comprehensive search for dark matter annihilation in dwarf galaxies}},  {\em
  \prd} {\bf 91} (Apr., 2015) 083535,
  [\href{http://xxx.lanl.gov/abs/1410.2242}{{arXiv:1410.2242}}].

\bibitem{strigari2010}
R.~{Essig}, N.~{Sehgal}, L.~E. {Strigari}, M.~{Geha}, and J.~D. {Simon}, {\it
  {Indirect dark matter detection limits from the ultrafaint Milky Way
  satellite Segue 1}},  {\em \prd} {\bf 82} (Dec., 2010) 123503,
  [\href{http://xxx.lanl.gov/abs/1007.4199}{{arXiv:1007.4199}}].

\bibitem{12}
A.~N. {Baushev}, S.~{Federici}, and M.~{Pohl}, {\it {Spectral analysis of the
  gamma-ray background near the dwarf Milky Way satellite Segue 1: Improved
  limits on the cross section of neutralino dark matter annihilation}},  {\em
  \prd} {\bf 86} (Sept., 2012) 063521,
  [\href{http://xxx.lanl.gov/abs/1205.3620}{{arXiv:1205.3620}}].

\bibitem{magic}
J.~{Aleksi{\'c}}, E.~A. {Alvarez}, L.~A. {Antonelli}, P.~{Antoranz},
  M.~{Asensio}, M.~{Backes}, J.~A. {Barrio}, D.~{Bastieri}, J.~{Becerra
  Gonz{\'a}lez}, W.~{Bednarek}, A.~{Berdyugin}, K.~{Berger}, E.~{Bernardini},
  A.~{Biland}, O.~{Blanch}, R.~K. {Bock}, A.~{Boller}, G.~{Bonnoli}, D.~{Borla
  Tridon}, I.~{Braun}, T.~{Bretz}, A.~{Ca{\~n}ellas}, E.~{Carmona},
  A.~{Carosi}, P.~{Colin}, E.~{Colombo}, J.~L. {Contreras}, J.~{Cortina},
  L.~{Cossio}, S.~{Covino}, F.~{Dazzi}, A.~{De Angelis}, E.~{De Cea del Pozo},
  B.~{De Lotto}, C.~{Delgado Mendez}, A.~{Diago Ortega}, M.~{Doert},
  A.~{Dom{\'{\i}}nguez}, D.~{Dominis Prester}, D.~{Dorner}, M.~{Doro},
  D.~{Elsaesser}, D.~{Ferenc}, M.~V. {Fonseca}, L.~{Font}, C.~{Fruck}, R.~J.
  {Garc{\'{\i}}a L{\'o}pez}, M.~{Garczarczyk}, D.~{Garrido}, G.~{Giavitto},
  N.~{Godinovi{\'c}}, D.~{Hadasch}, D.~{H{\"a}fner}, A.~{Herrero},
  D.~{Hildebrand}, D.~{H{\"o}hne-M{\"o}nch}, J.~{Hose}, D.~{Hrupec},
  B.~{Huber}, T.~{Jogler}, S.~{Klepser}, T.~{Kr{\"a}henb{\"u}hl}, J.~{Krause},
  A.~{La Barbera}, D.~{Lelas}, E.~{Leonardo}, E.~{Lindfors}, S.~{Lombardi},
  M.~{L{\'o}pez}, E.~{Lorenz}, M.~{Makariev}, G.~{Maneva}, N.~{Mankuzhiyil},
  K.~{Mannheim}, L.~{Maraschi}, M.~{Mariotti}, M.~{Mart{\'{\i}}nez},
  D.~{Mazin}, M.~{Meucci}, J.~M. {Miranda}, R.~{Mirzoyan}, H.~{Miyamoto},
  J.~{Mold{\'o}n}, A.~{Moralejo}, P.~{Munar-Androver}, D.~{Nieto},
  K.~{Nilsson}, R.~{Orito}, I.~{Oya}, S.~{Paiano}, D.~{Paneque}, R.~{Paoletti},
  S.~{Pardo}, J.~M. {Paredes}, S.~{Partini}, M.~{Pasanen}, F.~{Pauss}, M.~A.
  {Perez-Torres}, M.~{Persic}, L.~{Peruzzo}, M.~{Pilia}, J.~{Pochon},
  F.~{Prada}, P.~G. {Prada Moroni}, E.~{Prandini}, I.~{Puljak}, I.~{Reichardt},
  R.~{Reinthal}, W.~{Rhode}, M.~{Rib{\'o}}, J.~{Rico}, S.~{R{\"u}gamer},
  A.~{Saggion}, K.~{Saito}, T.~Y. {Saito}, M.~{Salvati}, K.~{Satalecka},
  V.~{Scalzotto}, V.~{Scapin}, C.~{Schultz}, T.~{Schweizer}, M.~{Shayduk},
  S.~N. {Shore}, A.~{Sillanp{\"a}{\"a}}, J.~{Sitarek}, D.~{Sobczynska},
  F.~{Spanier}, S.~{Spiro}, A.~{Stamerra}, B.~{Steinke}, J.~{Storz},
  N.~{Strah}, T.~{Suri{\'c}}, L.~{Takalo}, H.~{Takami}, F.~{Tavecchio},
  P.~{Temnikov}, T.~{Terzi{\'c}}, D.~{Tescaro}, M.~{Teshima}, M.~{Thom},
  O.~{Tibolla}, D.~F. {Torres}, A.~{Treves}, H.~{Vankov}, P.~{Vogler}, R.~M.
  {Wagner}, Q.~{Weitzel}, V.~{Zabalza}, F.~{Zandanel}, R.~{Zanin},
  M.~{Fornasa}, R.~{Essig}, N.~{Sehgal}, and L.~E. {Strigari}, {\it {Searches
  for dark matter annihilation signatures in the Segue 1 satellite galaxy with
  the MAGIC-I telescope}},  {\em \jcap} {\bf 6} (June, 2011) 35,
  [\href{http://xxx.lanl.gov/abs/1103.0477}{{arXiv:1103.0477}}].

\bibitem{segue12010}
P.~{Scott}, J.~{Conrad}, J.~{Edsj{\"o}}, L.~{Bergstr{\"o}m}, C.~{Farnier}, and
  Y.~{Akrami}, {\it {Direct constraints on minimal supersymmetry from Fermi-LAT
  observations of the dwarf galaxy Segue 1}},  {\em \jcap} {\bf 1} (Jan., 2010)
  31, [\href{http://xxx.lanl.gov/abs/0909.3300}{{arXiv:0909.3300}}].

\bibitem{frebel2014}
A.~{Frebel}, J.~D. {Simon}, and E.~N. {Kirby}, {\it {Segue 1: An Unevolved
  Fossil Galaxy from the Early Universe}},  {\em \apj} {\bf 786} (May, 2014)
  74, [\href{http://xxx.lanl.gov/abs/1403.6116}{{arXiv:1403.6116}}].

\bibitem{loeb2008}
T.~J. {Cox} and A.~{Loeb}, {\it {The collision between the Milky Way and
  Andromeda}},  {\em \mnras} {\bf 386} (May, 2008) 461--474,
  [\href{http://xxx.lanl.gov/abs/0705.1170}{{arXiv:0705.1170}}].

\bibitem{ahn2005}
K.~{Ahn} and E.~{Komatsu}, {\it {Cosmological lower bound on dark matter masses
  from the soft gamma-ray background}},  {\em \prd} {\bf 71} (Jan., 2005)
  021303, [\href{http://xxx.lanl.gov/abs/astro-ph/0412630}{{astro-ph/0412630}}].

\bibitem{klypin2013}
A.~{Klypin}, F.~{Prada}, G.~{Yepes}, S.~{Hess}, and S.~{Gottlober}, {\it {Halo
  Abundance Matching: accuracy and conditions for numerical convergence}},
  {\em ArXiv e-prints} (Oct., 2013)
  [\href{http://xxx.lanl.gov/abs/1310.3740}{{arXiv:1310.3740}}].

\end{thebibliography}

\providecommand{\href}[2]{#2}\begingroup\raggedright\endgroup

\end{document}